\documentclass[12pt]{iopart}

\expandafter\let\csname equation*\endcsname\relax

\expandafter\let\csname endequation*\endcsname\relax
\usepackage{xcolor}
\usepackage{mathtools}
\usepackage{wasysym}
\usepackage{tabularx}
\usepackage{hyperref}
\usepackage{cleveref}
\usepackage{graphicx}
\usepackage{adjustbox}
\usepackage{bm}
\usepackage{soul}
\usepackage{threeparttable}

\newcommand{\red}[1]{\color{black}{#1}\color{black}}

\newcommand{\green}[1]{\color{black}{#1}\color{black}}
\newcommand{\textout}[1]{}

\newcommand{\iu}{\mbox{i}}

\begin{document}

\title[Impact of the superconductors properties on haloscopes sensitivity]{Impact of the superconductors properties on the measurement sensitivity of resonant-based axion detectors}

\author{Andrea Alimenti $^{1}$,  Kostiantyn Torokhtii $^{1}$, Daniele Di Gioacchino $^{2}$, Claudio Gatti $^{2}$,  Enrico Silva $^{1,3}$ and Nicola Pompeo $^{1,3}$}

\address{$^{1}$ \quad Universit\`a degli Studi Roma Tre, Dipartimento di Ingegneria Industriale, Elettronica e Meccanica, Via Vito Volterra 62, 00146 Roma, Italy; \\
$^{2}$ \quad INFN, Laboratori Nazionali di Frascati, Frascati, Roma, Italy; \\
$^{3}$ \quad INFN, Sezione Roma Tre, Roma, Italy; }
\ead{andrea.alimenti@uniroma3.it}
\vspace{10pt}
\begin{indented}
\item[]Dec 2021
\end{indented}

\begin{abstract}
\red{Axions, hypothetical particles theorized to solve the strong CP-problem, are presently being considered as strong candidates as cold dark matter constituents. }
%
\red{The signal power of resonant-based axion detectors, known as haloscopes, is directly proportional to their quality factor $Q$. In this paper, the impact of the use of superconductors in the performances of the haloscopes is studied by evaluating the obtainable $Q$. }
In particular, the surface resistance $R_s$ of NbTi, Nb$_3$Sn, YBa$_2$Cu$_3$O$_{7-\delta}$ and FeSe$_{0.5}$Te$_{0.5}$ is computed in the frequency, magnetic field and temperature ranges of interest, starting from the measured vortex motion
\red{complex resistivity } and screening {length\red{s}} \red{of } these materials. From $R_s$ the quality factor $Q$ of a cylindrical haloscope with copper conical bases and superconductive lateral wall, operating with the TM$_{010}$ mode, is evaluated and used to perform a comparison of the performances of the different materials. Both YBa$_2$Cu$_3$O$_{7-\delta}$ and FeSe$_{0.5}$Te$_{0.5}$ are shown to \red{improve the measurement sensitivity by almost an order of magnitude } with respect to a whole Cu cavity, while NbTi is shown to be suitable only at lower frequencies ($<10$~GHz). Nb$_3$Sn can give an intermediate improvement in the whole spectrum of interest. 
\end{abstract}

%
%
%
%
%

\section{Introduction}
\red{The axion is a particle } associated to the dynamic field introduced by Peccei-Quinn in 1977 \cite{peccei1977cp,wilczek1978problem,weinberg1978new} to solve the `\textit{strong CP problem'}, i.e. the missing experimental observation of the \textit{CP} symmetry violation foreseen by the quantum chromodynamic theory. The characteristics of this particle make it particularly appealing also for the resolution of another unsolved physical problem: the nature \red{and existence } of \textout{the} dark matter. Thus, \red{in the last years } a \textout{large} \red{strong } interest \red{emerged } in the experimental detection of \textout{the} axions \textout{emerged in the last years}.


Due to the \red{small cross section} \textout{ for the interaction} of axions with the ordinary \red{baryonic } matter and \red{with } photons, the different detection approaches \textout{of this particle} exploit the Primakoff effect \cite{primakoff1951photo}, that is the conversion of axions in photons in presence of high magnetic fields $B$. The frequency of the produced photons $\gamma_{ph}$ is $\nu=E_{ph}/h$ with $E_{ph}$ the total energy of the particle (including the rest-mass energy and the kinetic contribution) and $h$ the Planck constant. 
\textout{Practically} \green{Thus, $E_{ph}=m_a c^2+\frac{1}{2}m_av^2$}, \textout{with} \red{where } $m_a$ \red{is } the mass of the axions, \green{ $c$ the speed of light in vacuum and $v$ the particle speed, which, for axions and compatible with the cold dark matter properties, is $<10^{-3}c$ \cite{tanabashi2018review}. Hence, in practice, $E_{ph}\approx m_a c^2$}. $m_a$ is constrained by astronomical and cosmological  considerations in the interval $1<m_a /(\mbox{\textmu eV})<10^{3}$ \cite{tanabashi2018review}.
Hence, ${2\times10^{8}\leq\nu/(\mbox{Hz})\leq2\times10^{11}}$. 

In 1983, P. Sikivie \cite{sikivie1983experimental} proposed an experimental test of the \red{existence of the } axions based on the use of resonant microwave cavities, known as haloscopes. When the axions enter the cavity, they can be converted into microwave photons by externally generated $B$ fields and in this way \red{they can be } detected \textout{through} \red{using } very low noise electronics \cite{braine2020extended}. The signal power $P_{a\rightarrow\gamma_{ph}}$ to be detected is shown to be \cite{sikivie1983experimental,braine2020extended}:

\begin{equation}\label{eqn:Paxions}
P_{a\rightarrow\gamma_{ph}}\propto\left(B^2VQ\right)\left(g_{a\gamma_{ph}}^2\rho_a\nu\right)\;,
\end{equation}
where $V$ is the volume of the cavity, $Q$ the quality factor, $g_{a\gamma}$ the axion-photon coupling and $\rho_a$ the local density of axions. \red{On the basis of existing experiments, one finds } $10^{-23}<P_{a\rightarrow\gamma_{ph}}/\mbox{(W)}<10^{-22}$. \textout{, based on existing experiments}. In \red{the proportionality statement~(\ref{eqn:Paxions})}, the terms in the right brackets do not depend on human will. Thus, the only change to improve the signal power, in order to reduce the integration time, consists in the optimization of the parameters in the left brackets \red{of (\ref{eqn:Paxions}).}

The high $Q$ needed to increase $P_{a\rightarrow\gamma_{ph}}$ pushes toward the study of new methods \textout{useful} to reduce \red{the cavity losses}.
The best results are obtained with either superconductive \cite{alesini2019galactic,ahn2020superconducting,golm2021thin} or dielectric resonant cavities \cite{alesini2021realization,alesini2020high}. Since in metallic cavities $Q\propto R_s^{-1}$, being $R_s$ the surface resistance of the conductive walls of the cavity, materials with low $R_s$ must be employed. Beyond copper, only superconductive materials can be explored. However, the choice is not obvious since \textout{it must be taken into account that} when superconductors (SC) are used at high frequencies and \red{in } high magnetic fields, they are driven in the mixed state \red{so that } large losses emerge due to the dissipative motion of \textout{the} fluxons. \red{Fluxons} \textout{The latter } are quanta of magnetic flux, of amplitude ${\Phi_0=h/(2e)\approx2.068\times10^{-15}}$~Wb, that penetrate \textout{the so called} type-II superconductors in presence of the high magnetic field levels needed \red{for the haloscopes } (i.e. of the order of some tesla). \textout{Thus,} \textout{since} In these conditions the $R_s$ of technological superconductors is dominated by the electrodynamics of fluxons. \red{The latter } must \red{then } be accurately characterized in order to assess the \textout{usability} \red{suitability } of these materials for the realization of highly sensitive haloscopes. 

Different superconductive haloscopes have been recently developed and tested. In \cite{alesini2019galactic} the $Q$-factor of copper and NbTi haloscopes (tuned at $\sim9.08$~GHz) have been measured, showing that \red{with a NbTi cavity } the  figure of merit $B^2Q$ at $\sim6$~T is $\sim5$ times larger \textout{with the NbTi cavity} than what reached with \textout{the} \red{a } Cu \textout{one} \red{cavity}. The first RE-Ba\textsubscript{2}Cu\textsubscript{3}O$_{7-\delta}$ (RE-BCO) cavity, with RE a rare earth element \red{such } as Y, was shown in \cite{ahn2020superconducting} to exhibit \red{$Q\approx3.25\times10^{5}$ } at 6.93~GHz and up to 8~T, which is more than 6 times larger than \textout{that obtained} with \red{a } Cu \red{cavity}. Recently \textout{also in} \cite{golm2021thin}, a different configuration \red{for a } \textout{of} RE-BCO cavity was presented and its performances compared with one coated with Nb$_3$Sn, both working at $\sim9$~GHz: the  RE-BCO haloscope reached $Q\sim7\times10^4$ at $\sim12$~T which was shown to be only 1.75 times \red{larger than } that obtained with the equivalent Cu \red{cavity}. Moreover, the $Q$-factor of the Nb$_3$Sn cavity showed a strong field dependence that apparently prevents the use of this material for this \red{kind of } application\red{s}.

In this paper, the most promising superconductors for resonant axion detectors, i.e. Nb$_3$Sn, NbTi, YBa$_2$Cu$_3$O$_{7-\delta}$ (YBCO) and, as a perspective, FeSe$_{0.5}$Te$_{0.5}$ (FeSeTe), \red{are considered. } \red{Their $R_s$ } is evaluated in the expected operative conditions of haloscopes, starting from the studies of the vortex motion properties performed on these materials. \green{In particular, the data used to perform the analysis on  Nb$_3$Sn and NbTi are taken respectively from \cite{alimenti2020microwave} and \cite{di2019microwave}, whereas those used to analyse the performances of YBCO and FeSeTe were measured specifically for this study. } From $R_s$, the figure{\red{s}} of merit $B^2Q$ of perspective haloscopes, based on the cylindrical geometry shown in \cite{alesini2019galactic,di2019microwave} and coated with these materials, \red{are } computed and compared.

\section{Physical background}\label{sec:Phys}
In this section the characteristics of the high frequency surface resistance $R_s$ of superconductors in the mixed state are introduced. First, the electrodynamics of \textout{the} vortex motion is presented and then its effects on $R_s$ \red{are } analysed \red{in terms of the field and the frequency dependences}.
\subsection{High frequency vortex motion}
\red{In } the microwave frequency range, \textout{the} fluxons are set in oscillation  around their equilibrium positions by the impinging electromagnetic (e.m.) wave. With low excitations and high frequencies (i.e. $>8$~GHz), as it is of interest in this case, the amplitude \red{displacement } of the vortices \red{from their equilibrium position } is generally smaller than the intervortex \red{spacing. Thus, } \textout{for this analysis } \red{in this regime fluxons can be assumed to be non-interacting with each other, } and a single-vortex motion approach can be followed to obtain the physical model \cite{coffey1991unified,gittleman1966radio,wu1995frequency,pompeo2008reliable,pompeo2020physics}. 

The force (per unit length) balance equation for a single fluxon is written down:

\begin{equation}\label{eqn:FbEquationFlux}
\bm J\times({\hat n} \Phi_0)+\bm F_{therm}=\eta \bm v+k_p \bm u\;,
\end{equation}
\textout{with} \red{where } $\bm J$ \red{is } the microwave-induced current density. \textout{which, interacting} \red{Its interaction} with the fluxons oriented along the ${\hat n}$ direction of the \textout{applied} magnetic field \red{$\bm B$}, \textout{is responsible for } \red{gives rise to } the driving Lorentz force that sets the fluxons \textout{themselves} in motion. $\bm F_{therm}$ is \textout{the} \red{a } stochastic \red{force on fluxons } \textout{contribution to the motion} given by \textout{the} thermal agitation, $\bm v$ is the \red{fluxon } velocity vector, $\bm u$ the displacement vector of the fluxon by its pinning centre, $\eta$ the viscous drag coefficient, and $k_p$ the elastic recall constant representing the pinning of the fluxon on its equilibrium position. 

At the temperatures of interest \red{for the haloscopes}, thermal creep effects can be neglected. Thus, by solving Equation~\eqref{eqn:FbEquationFlux} with respect to $\bm v$ and setting $\bm F_{therm}\rightarrow0$, one obtains the vortex-motion resistivity \cite{gittleman1966radio,pompeo2008reliable,pompeo2020physics}:

\begin{equation}\label{eqn:RhoGR}
\tilde\rho_{v}=\alpha\rho_{ff}\frac{1}{1-\iu \nu_p/\nu}\;,
\end{equation}
\textout{with} \red{where } $\rho_{ff}=\Phi_0 B/\eta$ represents the resistivity of \textout{the} completely free vortices, known as flux-flow resistivity, and $\alpha(\phi)=\sin^2(\phi)$ \cite{pompeo2015, pompeo2018analysis} \red{is } an adimensional correction coefficient taking into account the mutual orientation $\phi$ of the field and currents ($\phi=0$ when $\bm J\parallel\hat n$). Hence, $\tilde\rho_{v}$ is a complex quantity characterized by \textout{the presence of} the characteristic frequency $2\pi\nu_p=k_p/\eta$ which separates an elastic low-frequency vortex motion regime from \red{a highly } dissipative motion regime at higher frequencies. For $\nu\gg\nu_p$, $\tilde\rho_v\rightarrow\rho_{ff}$. \red{The latter } can be a rather large quantity \red{at high fields}, being a fraction $\rho_{ff}\sim\rho_nB/B_{c2}$ of the normal state resistivity $\rho_n$, \red{with } $B_{c2}$ the upper critical field of the material \cite{tinkham2004introduction,bardeen1965theory}. 

In anisotropic materials (e.g. YBCO and FeSeTe) the vortex motion parameters exhibit an angular dependence given by the different properties of the material along the crystallographic directions. In case of uniaxial anisotropy, \red{resulting } by the layered structure of some superconductors, an anisotropic effective electron mass can be measured. This brings to the definition of the anisotropy coefficient $m_c=\gamma^2m_{ab}$, 
where $m_{c}$ and $m_{ab}$ are the effective masses along the $c$-axis \red{(the crystallographic axis of anisotropy) } and the $ab$-planes \textout{being} \red{($a$ and $b$ are the crystallographic axes normal to the $c$-axis) } respectively, and $\gamma$ \red{is } the intrinsic material anisotropy coefficient. The anisotropy on the vortex motion parameters, for high-$\kappa$ SC in the London limit \cite{tinkham2004introduction}, is then derived following the Blatter-Geshkenbein-Larkin (BGL) scaling theory \cite{blatter1992isotropic} which states that for a field and angle dependent observable $\mathcal{Q}(H, \theta)$, its $H$ and $\theta$ \textout{being} ($\theta$ \red{is } the angle between $\bm H$ and the $c$-axis) dependences can be combined in an effective field $H\epsilon(\theta)$:
\begin{subequations}
\begin{eqnarray}
\mathcal{Q}(H, \theta)&=&s_{\mathcal{Q}}(\theta)\mathcal{Q}(H\epsilon(\theta))\\
\epsilon(\theta)&=&(\gamma^{-2}\sin^2\theta+\cos^2\theta)^{1/2}
\end{eqnarray}
\label{eq:scaling}
\end{subequations}
where $s_{\mathcal{Q}}(\theta)$ is an a-priori known factor which depends on the particular $\mathcal{Q}(H, \theta)$: $s_{\eta}(\theta)=\epsilon(\theta)^{-1}$ and $s_{k_p}(\theta)=\epsilon(\theta)^{-1}$ \cite{pompeo2018analysis,pompeo2020intrinsic,pompeo2020pinning}. However, it should be \red{noted } that the scaling of $k_p$ with the BGL theory can be successfully performed only when isotropic pinning centres \red{are present } \green{as discussed and experimentally shown in \cite{pompeo2020intrinsic}}.
For this reason in the paper only samples without the addition of directional pinning centres are analysed.

Finally, when $\bm B\parallel\bm J$, i.e. $\phi=0$, $\alpha=0$ since the driving force acting on the fluxon $\bm J\times(\hat n \Phi_0)=0$. However, \red{since fluxons are magnetic structures,  inside the superconductor the direction of the magnetic flux density field $\bm B$ can differ from the direction of the outer static field $\bm H$, hence $\bm B\neq\mu_0\bm H$.  Indeed }  also with nominally $\bm H\parallel\bm J$ it was experimentally observed that $\tilde\rho_v\neq0$ \cite{di2019microwave}. 
This can be explained considering that the finite line tension ${\varepsilon_\ell=\Phi_0^2/(4\pi\mu_0\lambda^2)}$ of the fluxon lines, \green{with $\lambda$ the London penetration depth}, allows them to locally deviate from the $\bm{H}$ orientation by an angle $\phi_{loc}$, so that this finite angle between $\bm J$ and the flux lines yields $\alpha\neq0$. 
Indeed, vortex lines deform under the effects of the attractive action of pinning centres, of thermal fluctuations and of the Lorentz force exerted by the applied currents. The latter is well known to produce helicoidal current vortices, with the corresponding instabilities in \red{DC } regimes, and more generally flux cutting phenomena \cite{campbell2011a}. In the present situation we are considering vanishingly small, high frequency currents whose contribution to vortex deformation and tilting with respect to their nominal orientation can be neglected. 
On the other hand, the effect of pinning centres, here considered to be point-like, and thermal fluctuations, can be evaluated in a simplified way within the collective pinning theory \cite{blatter1994}. 
Thanks to their finite elasticity, vortices accommodate on the underlying pinning landscape to minimize the total elastic \textout{plus} \red{and } pinning energy. In so doing, they displace by an amplitude given by the interaction length with pins ($\sim\xi$) over the so called collective pinning length $L_c$. \red{The latter } identifies the length of the vortex segment which is independently pinned with respect to the other portions of the vortex. In a simple picture, this accommodation produces a tilting angle $\phi_{loc,el}\simeq{\rm{arctan}}(\xi/L_c)$ between the flux line and the nominal $\bm{H}$ field orientation.
The length $L_c$ is connected to the vortex collective pinning energy $U_c\simeq \varepsilon_\ell \xi^2 /L_c$. 
\textout{On the other hand,} \red{Moreover, } the average amplitude of the thermal fluctuations $u_{th}$ can be estimated \red{by } equating the thermal energy per unit length $k_B T/L_c$ to the elastic energy per unit length $(1/2)\varepsilon_\ell u^2_{th}$ due to these deformations. 
Thus, the tilting angle due to thermal fluctuations is $\phi_{loc,th}\simeq{\rm{arctan}}(u_{th}/L_c)$. 
Since the two mechanisms are independent, a combined tilting angle can be computed $\phi_{loc}\simeq{\rm{arctan}}(\sqrt{u_{th}^2+\xi^2}/L_c)$. \green{The procedure used to estimate $\phi_{loc}$ from measured parameters will be detailed in Section~\ref{sec:Materials}.}

This evaluation will prove useful in the following.

\subsection{Surface impedance in the mixed state}
The macroscopic material property that describes the electrodynamic response of (super)conductors is the surface impedance $Z_s$ \cite{collin2007foundations}:

\begin{equation}
Z_s\coloneqq E_{\parallel}/H_{\parallel}=R_s+\iu X_s\;,
\end{equation}
\textout{with} \red{where } $E_{\parallel}$ and $H_{\parallel}$ \red{are } the electric and magnetic field component parallel to the surface of the material, \red{respectively, and } $R_s\coloneqq\mbox{Re } Z_s$ and $X_s\coloneqq\mbox{Im } Z_s$. 

In the local limit for bulk good conductors, i.e. when the thickness $d$ of the material is much larger than the screening characteristic lengths $d\gg\mbox{min}(\delta,\lambda)$, with $\delta$ the skin penetration depth and $\lambda$ the London penetration depth, from classical electromagnetism it can be shown that  \cite{collin2007foundations}:

\begin{equation}\label{eqn:Zbulk}
Z_s=\sqrt{\iu \omega \mu_0 \tilde\rho}\;,
\end{equation}
where $\omega=2\pi\nu$ is the angular frequency of the impinging e.m. field, $\mu_0$ is the vacuum magnetic permeability and $\tilde\rho$ the resistivity of the material which, in the case of superconductors, is a complex quantity. 
The resistivity $\tilde\rho$ describes all the resistive and reactive phenomena in the material. Thus, for superconductors in the mixed state the contribution of both \textout{families of} \red{superconducting and normal } charge carriers, \textout{(i.e. superconductive and normal) and} \red{as well as } vortex motion, must be taken into account in $\tilde\rho$. The interplay between the various contributions to $\tilde\rho$ \textout{in the mixed state}, for small displacements $\bm u$ of the fluxons, yields an overall resistivity which can be written as \cite{coffey1991unified}:

\begin{equation}\label{eqn:rhocompl}
\tilde\rho=\frac{\tilde\rho_v+\iu/\sigma_2}{1+\iu\sigma_1/\sigma_2}\;,
\end{equation}
where $\sigma_{2f}\coloneqq\sigma_1-\iu\sigma_2$ is the so called two-fluid conductivity of superconductors \cite{tinkham2004introduction}. \red{This conductivity } \textout{ which, at the frequency of interest,} can be described as the parallel between the real conductivity of the quasi-particles $\sigma_1$ \textout{$\approx\sigma_{QP}$} and the inductive behaviour of the superfluid ${\sigma_2\approx1/(\omega\mu_0\lambda^2)}$. 
Taking into account the low operative temperatures of haloscopes, since $\sigma_{1}\propto x_n$ with $x_n\approx(T/T_c)^\beta$ (with $\beta=4$ and $\beta=2$ for low-$T_c$ and high-$T_c$ SC, respectively), \red{$\sigma_1/\sigma_2\ll 1$. Hence, }
Equation~\eqref{eqn:rhocompl} can be approximated to $\tilde\rho\approx\tilde\rho_v+\iu/\sigma_2$ which yields:

\begin{equation}\label{eqn:ZbulkSup}
Z_s\approx\sqrt{\omega\mu_0\left(-\frac{1}{\sigma_2}+\iu\tilde\rho_v\right)}\;.
\end{equation} 
Equation~(\ref{eqn:ZbulkSup}) will be used in the following to evaluate and compare the materials surface resistance $R_s$ starting from the measurements of the vortex motion parameters, $\rho_{ff}$ and $\nu_p$, which define $\tilde\rho_{v}$ in Equation~\eqref{eqn:RhoGR}, and $\lambda$.

\section{Experimental section}
In this section the surface resistance of NbTi, Nb\textsubscript{3}Sn, YBCO \red{and } FeSeTe is evaluated starting from the measurements of the vortex motion parameters. Then, the perspective performances of haloscopes coated with these materials are analysed. 

To this purpose, we first show the structure of a typical haloscope taking as \red{a } reference \textout{that} \red{the haloscope } presented in \cite{alesini2019galactic} in order to highlight the dependence of $Q$ on the $R_s$ of the coating layer. Then, the  properties \red{of the superconductors } needed for the evaluation of $R_s$ are collected from existing works.
\textout{, and } Finally\red{, using the computed $R_s$ }\textout{from them} the sensitivity of haloscopes is \red{evaluated } to assess which of the \textout{se} SCs \red{under scrutiny } would give the best performances.

\subsection{Haloscope design}\label{sec:Haloscope}
\begin{figure}
\centering
\includegraphics[width=0.4\textwidth]{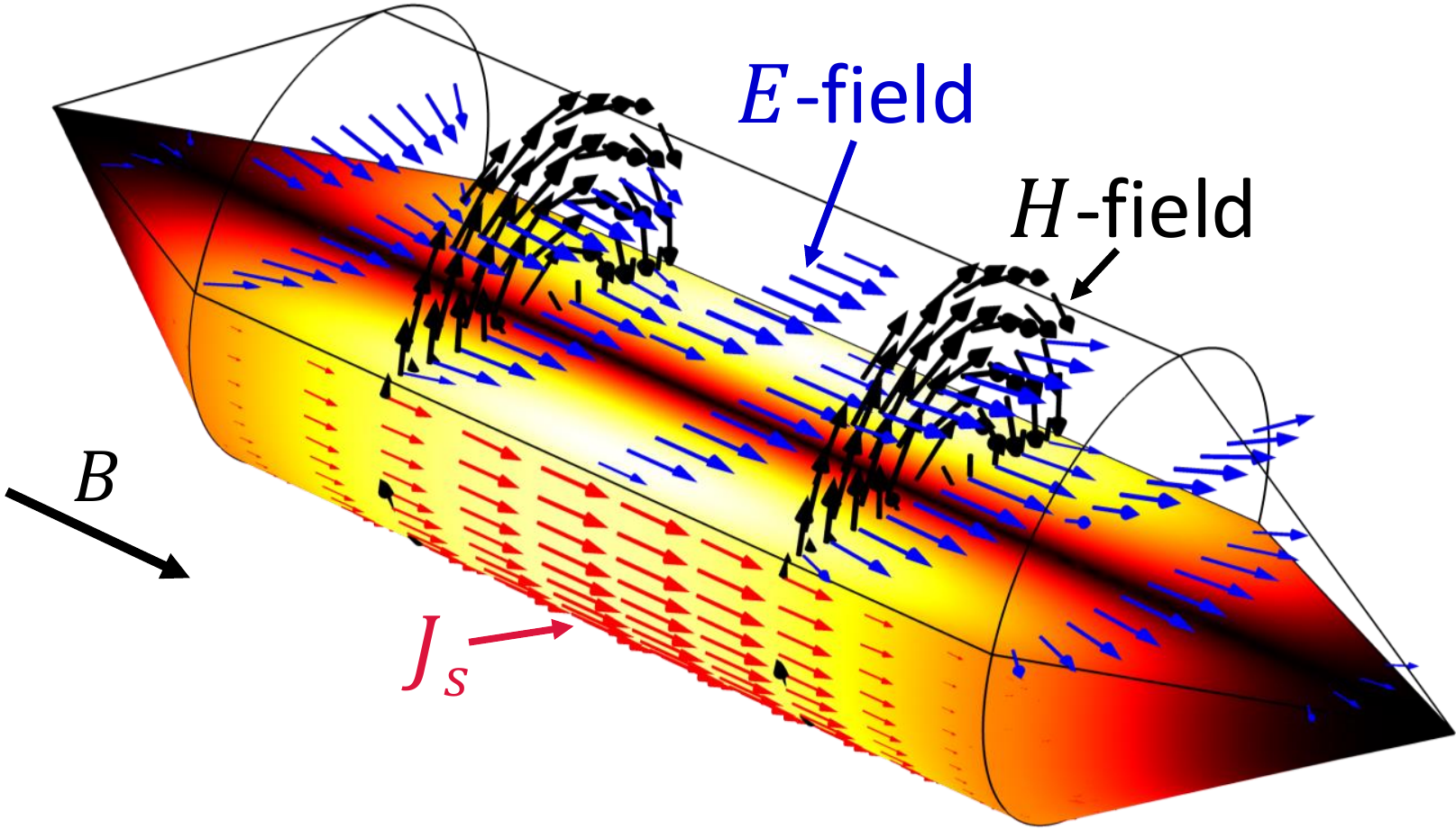}
\caption{\red{Electromagnetic simulation of the haloscope excited with the TM\textsubscript{010} mode. The distributions of the radio frequency electric $E$-field (in blue), the magnetic $H$-field (in black) and the surface current density $J_s$ (in red) are shown. The direction of the externally applied static $B$ field is shown to be parallel to the longitudinal axis of the cavity.}}
\label{fig:cavity}
\end{figure}

In this work, the superconductive haloscope presented in \cite{alesini2019galactic} within the QUAX experiment is used as \red{a } reference. \textout{for the comparison} The cavity has cylindrical shape ($\diameter=26.1$~mm) with conical bases and it is designed to operate with the applied magnetic field oriented along the main \red{axis } of the cavity. In order to facilitate the penetration of the magnetic field, the conical caps are made of copper while the cylindrical SC body is divided in two halves \textout{and} \red{by } a copper thin (30~\textmu m) layer \textout{placed between them} to break the \textout{SC} \red{superconducting } screening currents. The \red{transverse magnetic } TM\textsubscript{010} mode, which is usually employed for Primakoff axion detection in conjunction with a static magnetic field orientation parallel to the cavity axis, induces longitudinal currents on the lateral walls (thus \red{$\bm H \parallel \bm J$}). \textout{and} \red{The haloscope here considered } resonates at $\nu_0\sim9.08$~GHz at 4~K \cite{alesini2019galactic}. \red{The subscript indices of `TM\textsubscript{010}' indicate the number of the peaks of the radio frequency $H$-field met along the three directions of a cylindrical coordinate system: the first index `0' indicates that the $H$-field is constant along the azimuth direction, the second index `1' indicates that the $H$-field exhibits a peak along the radial direction and the last index `0' that the $H$-field is constant along the axial direction \red{(a sketch of the radio frequency fields and currents is reported in Figure~\ref{fig:cavity})}.}

Due to the different materials used for the inner coating of the cavity elements, the overall quality factor is given by:

\begin{equation}\label{eqn:Qhaloscope}
\frac{1}{Q}=\frac{R_{s,cyl}}{G_{cyl}}+\frac{R_{s,cones}}{G_{cones}}\;,
\end{equation}
\textout{with,} \red{where } for this cavity, $G_{cyl}\simeq482\;\Omega$ and $G_{cones}\simeq6270\;\Omega$ \red{are } the geometrical factors of the cylindrical body and conical caps, respectively, and $R_{s,cyl}$, $R_{s,cones}$ their surface resistances.  At 9.08~GHz and 4~K, in the anomalous skin depth regime \cite{reuter1948theory}, the surface resistance of copper is $R_{s,Cu}\simeq4.9\;\mbox{m}\Omega$. Thus, a cavity with this geometry entirely made of copper would have $Q_{min}\simeq9.1\times10^4$ at 9~GHz. This is the lower limit given by copper \textout{and} that we expect to be outperformed by the use of superconductive coatings. Whereas, in the ideal case for which $R_{s,cyl}/G_{cyl}$ can be neglected, the upper limit $Q_{max}\simeq1.3\times10^6$ could be reached with this geometry. Thus, the highest performance increase in the $B^2Q$ figure of merit reachable with this haloscope geometry is $Q_{max}/Q_{min}=G_{cones}/G_{cyl}+1\approx14$.

\subsection{Materials under investigation}\label{sec:Materials}
The perspective application of four different superconductive materials to the realization of haloscopes is here investigated. 

In Table~\ref{tab:samples} the list of the samples, whose vortex motion parameter were measured in high magnetic fields, their main characteristics, and references where further information can be found, are reported.
\begin{center}
\begin{table}[h!!!!]
\caption{List of the samples under investigation. 
}\label{tab:samples}
\begin{threeparttable}
\begin{tabular}{|c|c|c|c|c|} 
\hline
Material & NbTi & Nb\textsubscript{3}Sn & YBa\textsubscript{2}Cu\textsubscript{3}O\textsubscript{$7-\delta$} & FeSe\textsubscript{0.5}Te\textsubscript{0.5} \\ 
\hline\hline
Thickness (nm) & $(3.5\pm0.5)\times10^3$ & bulk & $80\pm5$ &$240\pm15$\\
\hline
Growing tech.& RF sputtering & HIP\tnote{1} & CSD\tnote{2} & PLD\tnote{3}\\
\hline
Substrate & Cu & -- & LaAlO\textsubscript{3} &CaF\textsubscript{2} \\
\hline
$T_c$ (K)&$8.0\pm0.2$&$18.0\pm0.2$&$89.9\pm0.5$&$18.0\pm0.2$\\
\hline
Ref.&\cite{alesini2019galactic}&\cite{flukiger2017variation,alimenti2020microwave} &\cite{pinto2020chemical}&\cite{braccini2013highly}\\
\hline
\end{tabular}
\begin{tablenotes}
\red{\item[1] High Isostatic Pressure
\item[2] Chemical Solution Deposition
\item[3] Pulsed Laser Deposition}
\end{tablenotes}
\end{threeparttable}
\end{table}
\end{center}

\textout{Among these,} The microwave vortex motion properties of NbTi were characterized through the study of the performance of the NbTi coated haloscope analysed in \cite{alesini2019galactic}. The Nb$_3$Sn, YBCO and FeSeTe samples were studied through the measurement technique based on the use of the dielectric loaded resonator described and optimized in \cite{alimenti2019challenging,torokhtii2021optimization,torokhtii2018q,alimenti2021surface}. In particular the analysis of the vortex motion parameters in Nb\textsubscript{3}Sn was reported in \cite{alimenti2020microwave,alimenti2019surface}. The microwave surface impedance analysis of FeSeTe was detailed in \cite{pompeo2020pinning,pompeo2021pinning,pompeo2020microwave}, while the YBCO measurements in high magnetic fields were presented in \cite{AlimentiASC2020}.

For the comparison we choose to work in the same operative condition of the cavity analysed in \cite{alesini2019galactic}: $T=4$~K, $\mu_0H=5$~T parallel to the SC surface, and with the selected resonant mode inducing microwave currents $\bm J\parallel \bm H$.
Actually, the latter samples (Nb$_3$Sn, YBCO and FeSeTe) were characterized with a different configuration of currents and field orientations ($\bm J\perp\bm H$ with $\bm H \parallel$ $c$-axis for anisotropic SC) with respect to the haloscope here considered. 
Hence, the vortex motion resistivity $\tilde\rho$ must be properly evaluated in the configuration of interest. 
First, the $\alpha$ coefficient must be evaluated, see Equation~\eqref{eqn:RhoGR}. This is done through the first order model proposed in Section \ref{sec:Phys}.\textout{, in which }
The \textout{needed} collective pinning length $L_c=\sqrt{2\varepsilon_\ell/k_p}$ is evaluated from the experimentally measured $k_p$ by equating the pinning energy $U_c/L_c$ to the pinning energy (both per unit length) expressed in terms of $k_p$, $k_p \xi^2/2$. \green{The data used to estimate $L_c$ and $\varepsilon_\ell$, i.e. $\lambda$ and $k_p$, are reported in Table~\ref{tab:Vpar}. }


Second, for both YBCO and FeSeTe the intrinsic material anisotropy must be taken into account (see Equation~\eqref{eq:scaling}), 
taking $\gamma$ from \cite{pompeo2020intrinsic} and \cite{pompeo2021pinning} for YBCO and FeSeTe, respectively. 
The anisotropy enters also the evaluation of the $\alpha$ coefficient which, in the geometry here considered, is influenced by the anisotropy which enhances tilting due to thermal fluctuations and reduces the one due to the elastic accommodation to pins. 
It is worth noting here that \textout{the} FeSeTe \textout{superconductor} is a multiband superconductor. \red{Indeed, it exhibits }\textout{an} intricate anisotropy properties which are actually different among the various superconducting quantities $\lambda$, $\xi$, $B_{c1}$ and $B_{c2}$ \cite{bendele2010anisotropic}. \red{Moreover, }$\rho_{ff}$ shows \red{an } unusual \textout{properties such its} field dependence \cite{okada2015exceptional,pompeo2020pinning,pompeo2021pinning,pompeo2020microwave}. Hence a direct experimental study of FeSeTe in the same configuration involved in the considered haloscope design would be particularly recommended.

In addition, \green{for the sake of simplicity, } \red{we neglect the increased pinning when $\bm H \parallel ab$ planes (the so called ``intrinsic pinning''), } \textout{another source of anisotropy,} connected to the layered structure of materials like YBCO. \green{Moreover, it must be noted the in engineered materials the effects of the intrinsic pinning are essentially hidden by those of the artificial pinning centres \cite{pompeo2020intrinsic}}. \textout{ and manifesting as an increased pinning when $\bm H \parallel$ $ab$ planes (the so called ``intrinsic pinning'') is here neglected.}

\begin{figure}[htbp]
\centering
\includegraphics[width=1\columnwidth]{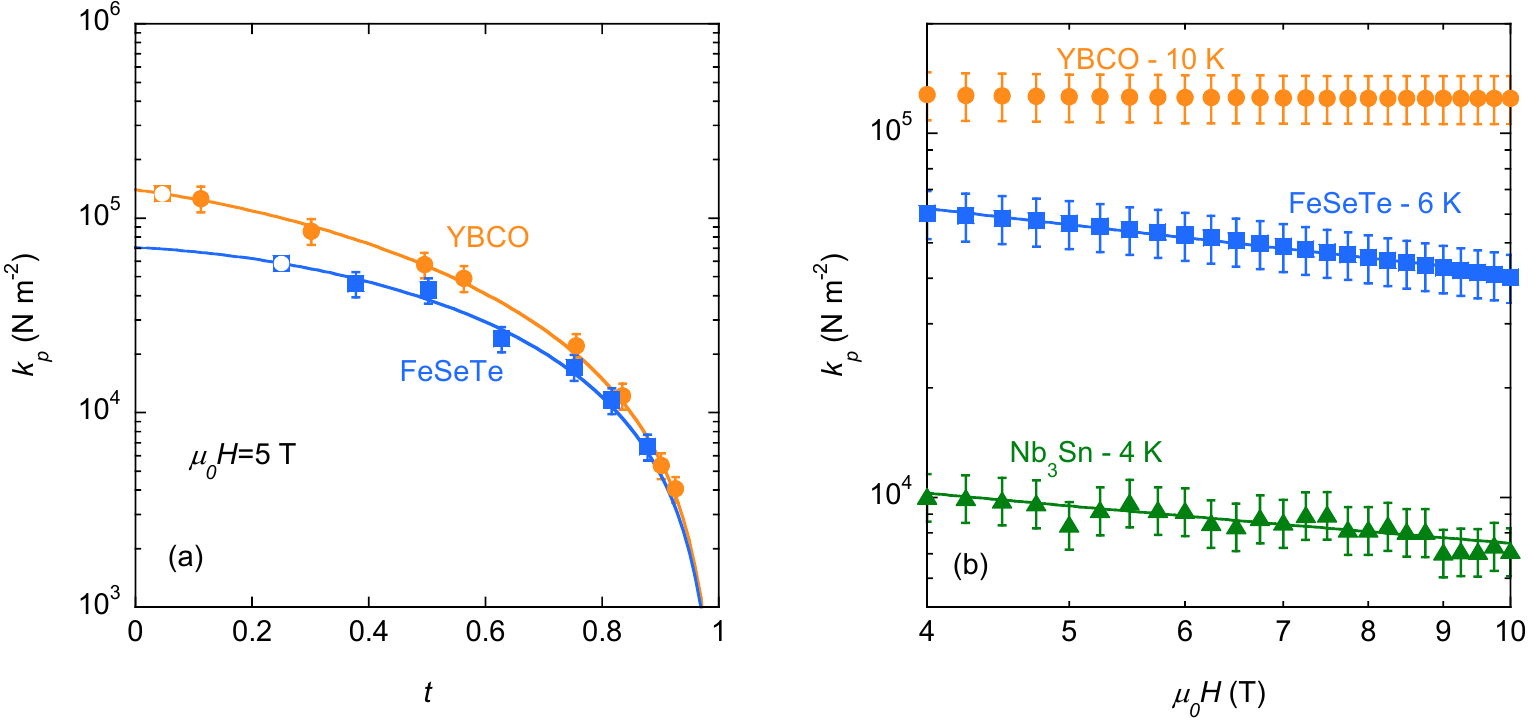}
\caption{(a) pinning constant $k_p$ measured on YBCO (orange circles) and FeSeTe (blue squares) in zero field cooling condition (ZFC) at 5~T as a function of the reduced temperature $t=T/T_c$. The fit of the data is performed with the model developed in \cite{golosovsky1996high}: ${k_p(t)=k_p(0)(1-t)^{4/3}(1+t)^2e^{(-T/T_0)}}$. The best fit parameters are: $k_p(0)=1.40\times10^5$~N~m\textsuperscript{-2} and $T_0=56$~K for YBCO, ${k_p(0)=7.07\times10^4}$~N~m\textsuperscript{-2} and $T_0=T_c$ for FeSeTe. The empty square symbols represent the points extrapolated from the fit at 4~K (which will be used in the next elaborations). (b) Field dependence of $k_p$ in Nb$_3$Sn (green triangle), YBCO and FeSeTe samples measured at the lower temperatures. $k_p(H)$ in Nb\textsubscript{3}Sn and FeSeTe is shown to be $\propto H^{-0.5}$.
}
\label{fig:KpLayout}
\end{figure}

\begin{figure}[htbp]
\centering
\includegraphics[width=0.5\columnwidth]{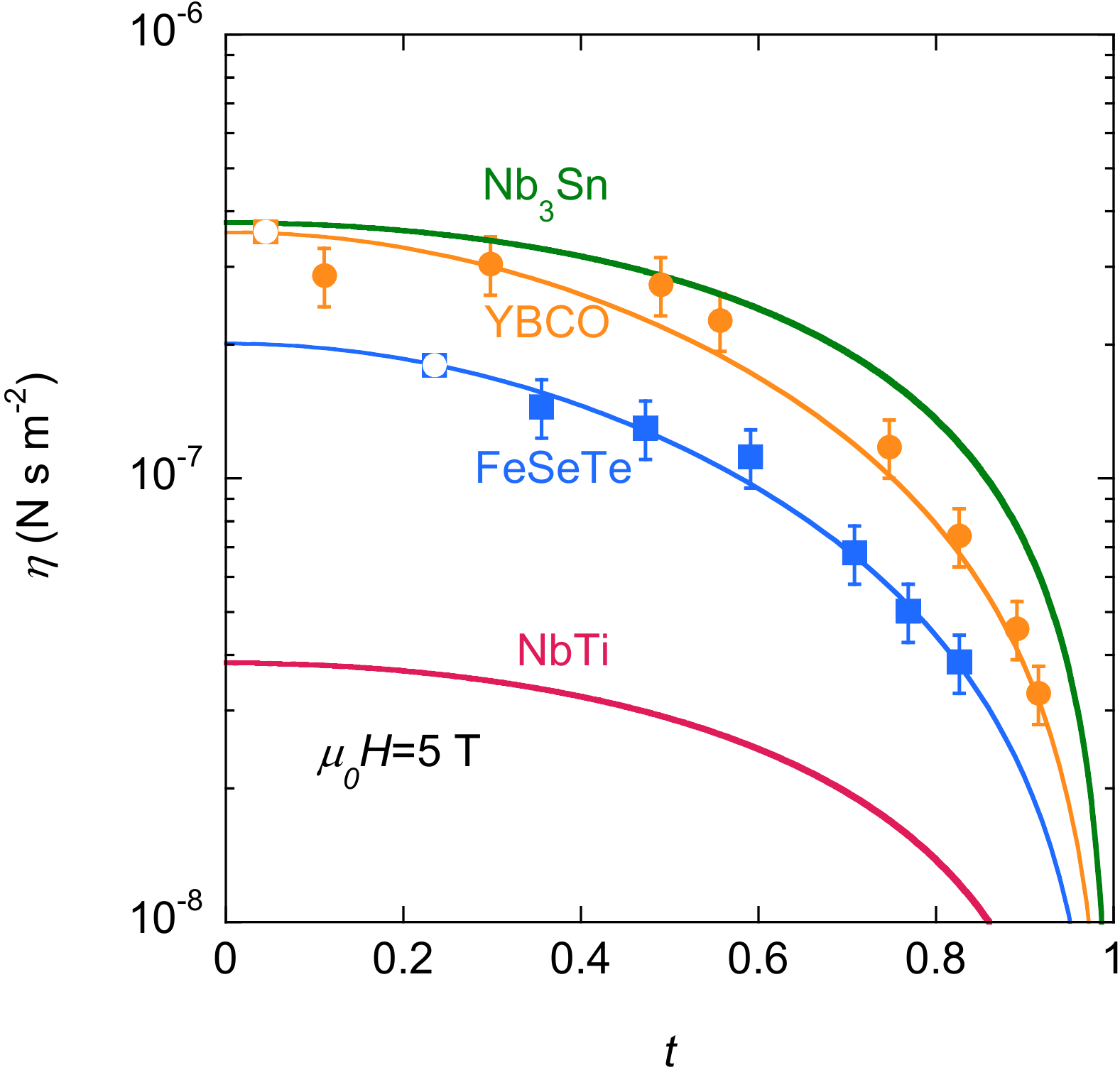}
\caption{Viscous drag coefficient $\eta$ measured on YBCO (orange circles) and FeSeTe (blue squares) at 5~T in ZFC as a function of the reduced temperature $t$. The measured data are approximated with $\eta(t)=\eta(0)(1-t^2)/(1+t^2)$ \cite{golosovsky1996high}, with $\eta(0)=3.59\times10^{-7}$~N~s~m\textsuperscript{-2} for YBCO and ${\eta(0)=2.02\times10^{-7}}$~N~s~m\textsuperscript{-2} for the FeSeTe sample. The empty square symbols represent the points extrapolated from the fit at 4~K (which will be used in the next elaborations). The green and red curves \textout{are the} \red{represent } $\eta(t)$ obtained \red{by } considering $\eta(t)=\Phi_0B_{c2}(t)/\rho_{n}$  (from $\rho_{ff}=\rho_nB/B_{c2}$ and $\eta=\Phi_0B/\rho_{ff}$) for Nb$_3$Sn and NbTi respectively. } 
\label{fig:etaVsT}
\end{figure}


Finally, the $T$ and $H$ range of interest must be considered. 
Since measurements at $T=4$~K and $\mu_0H=5$~T are available only for Nb$_3$Sn, for YBCO and FeSeTe the low temperature data must be extrapolated from currently existing measurements. However, the temperature dependences of both $\eta$ and $k_p$ derive from those of $\lambda$ and \red{of } the coherence length $\xi$ \cite{golosovsky1996high}: hence, at low temperature these parameters saturate similarly to $\lambda$ and $\xi$ \cite{tinkham2004introduction}. Thus, negligible uncertainties associated to the model choice are introduced in the mentioned extrapolation \cite{golosovsky1996high}.

 The pinning frequency is computed from $k_p$ and $\eta$ as $2\pi\nu_p=k_p/\eta$. 
In Figure~\ref{fig:KpLayout}(a) the pinning constant $k_p(t)$, with $t=T/T_c$, measured at 5~T on YBCO and FeSeTe is shown. The measurements are obtained in zero field cooling (ZFC) condition \red{by } sweeping the magnetic field at fixed temperature values. The values at 4~K are extrapolated from these measurements through the empirical models shown in \cite{golosovsky1996high} and reported in the caption of Figure~\ref{fig:KpLayout}. The field dependence of $k_p$ is shown in Figure~\ref{fig:KpLayout}(b). Here also the Nb\textsubscript{3}Sn measurement, performed at 4~K (thus for it no extrapolation is needed), is reported. From the \textout{shown} \red{behavior of }  $k_p(H)$ on the different materials, different pinning regimes can be recognized. In YBCO, $k_p$ is field independent at 10~K and up to 10~T, indicating \textout{that it is in} a pinning regime where fluxons are strongly pinned independently one from another.
On the contrary, \textout{both in} FeSeTe and Nb\textsubscript{3}Sn \red{display} a $k_p\propto H^{-0.5}$ dependence, typical of the collective pinning regime (\red{bundles } of fluxons pinned on multiple pinning centres).
\textout{is well distinguishable cite golosovsky1996high.}
These \red{field } dependences \textout{to the field} will be considered in the final performance analysis presented.

For what concerns $\eta=\Phi_0B/\rho_{ff}$, in Figure~\ref{fig:etaVsT} measurements at 5~T on YBCO and FeSeTe are reported. Also in this case, from the model shown in \cite{golosovsky1996high}, the value at 4~K is extrapolated. Resorting to the conventional Bardeen-Stephen behaviour \cite{bardeen1965theory} exhibited by Nb\textsubscript{3}Sn \cite{alimenti2020microwave} and NbTi, the flux-flow resistivity is simply obtained from the normal state resistivity $\rho_{n}$ as $\rho_{ff}=\rho_nB/B_{c2}$ with $B_{c2}\approx B_{c2}(0)(1-t^2)$. We take $\rho_{n}=14.8$~\textmu$\Omega$~cm and $B_{c2}=27$~T for Nb\textsubscript{3}Sn \cite{alimenti2020microwave} and $\rho_n=70$~\textmu$\Omega$~cm, $B_{c2}=13$~T for NbTi \cite{alesini2019galactic}.

All the parameters of interest for the different materials are reported in  Table~\ref{tab:Vpar}.

\begin{table}
\caption{Measured parameters of the SCs of interest used for the evaluation of $R_s$.}\label{tab:Vpar}
\begin{adjustbox}{width=\columnwidth}
\small
\begin{tabular}{|c|c|c|c|c|} 
\hline
& NbTi & Nb$_3$Sn & YBCO&FeSeTe \\ 
\hline\hline
$k_p$(N~m\textsuperscript{-2}) at 4~K, 5~T& -- & $(9.9\pm1.5)\times10^3$ & $(1.3\pm0.2)\times10^5$ &$(5.9\pm0.9)\times10^{4}$\\
\hline
pinning regime& single & collective & single &collective\\ 
\hline
$\eta$(N~s~m\textsuperscript{-2}) at 4~K, 5~T& -- & -- & $(3.6\pm0.5)\times10^{-7}$& $(1.8\pm0.3)\times10^{-7}$\\ 
\hline
$\rho_n$(\textmu$\Omega$~cm) &$70\pm2$ & $14.8\pm0.2$ & -- &--\\
\hline
$B_{c2}(0)$(T) &$13.0\pm0.5$ & $27.0\pm0.5$ & -- &--\\
\hline
$\nu_p$(GHz) at 4~K, 5~T& $44\pm7$ & $4.2\pm0.6$ & $59\pm9$ & $52\pm8$\\
\hline
meas. geometry & $\bm H \parallel \bm J$ & $\bm H \perp \bm J$ & \multicolumn{2}{|c|}{$\bm H \parallel$ $c$ axis, $\bm H \perp \bm J$}\\
\hline
$\gamma$& 1 & 1 & $5.3\pm0.7$ & $1.8\pm0.2$\\ 
\hline
$\lambda(0)$(nm)&$270\pm20$ \cite{di2019microwave,benvenuti91} & ${160\pm20}$ \cite{posen2014advances} & ${150\pm10}$ \cite{sonier1994new,tallon1995plane}&$520\pm50$ \cite{okada2015exceptional}\\ 
\hline
\end{tabular}
\end{adjustbox}
\end{table}

\section{Results and discussion}
From the data collected in Table~\ref{tab:Vpar}, the surface resistance $R_s$ of the selected SCs is evaluated to compare the figure of merit $B^2Q$ of haloscopes (\textout{having} \red{with } the geometry shown in \cite{alesini2019galactic}) coated with different materials. In particular, Equation~\eqref{eqn:RhoGR} is computed with the data in Table~\ref{tab:Vpar} taking into account the $\alpha$ coefficient, the anisotropy $\gamma$ for YBCO and FeSeTe, the collective pinning dependences in Nb$_3$Sn and FeSeTe and the pair breaking effects on the superfluid $\sigma_2$ and, for NbTi, also on $\sigma_1$. From this, the complex resistivity $\tilde\rho$ from Equation~\eqref{eqn:rhocompl} and $R_s$ from Equation~\eqref{eqn:Zbulk} are obtained. Finally, $Q$ of the haloscope is computed using Equation~\eqref{eqn:Qhaloscope}.

\begin{figure}[htbp]
\centering
\includegraphics[width=0.45\textwidth]{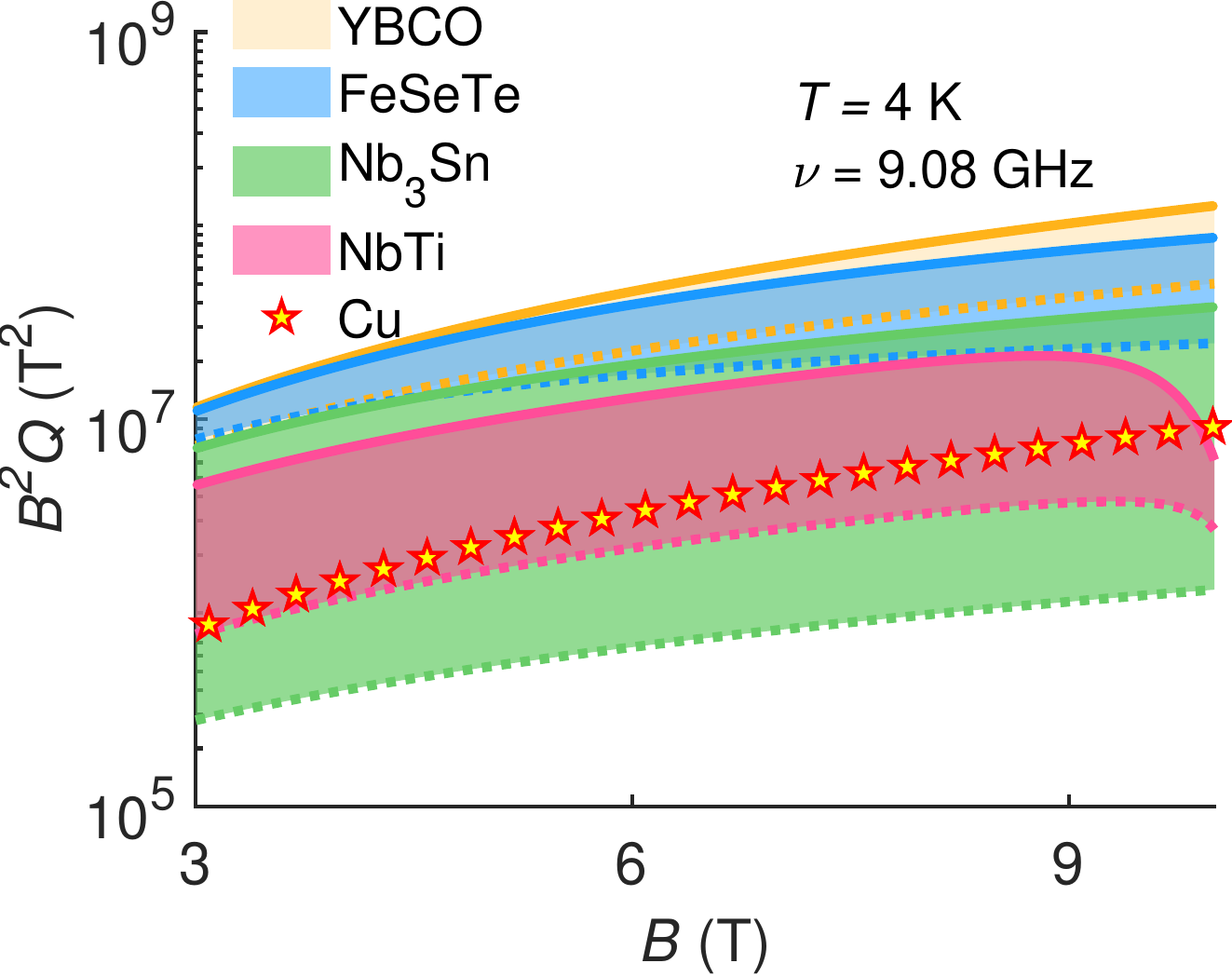}
\caption{Comparison of the $B^2Q$ factors evaluated at 4~K and 9.08~GHz as a function of $B$. The dotted lines represent the worst lower limit obtained in the $\bm B\perp\bm J$ configuration. The continuous lines represent a more realistic situation for which the $\alpha$ angular coefficient is taken into account as explained in the text. The ``star” symbols represent the $B^2Q$ values obtained with a Cu cavity, the yellow area the values obtainable with YBCO, the light blue area those with FeSeTe, the green with Nb$_3$Sn and the red with NbTi.}
\label{fig:Q2B_VsB}
\end{figure}

The performances comparison is based on the figure of merit $B^2Q$. In Figure~\ref{fig:Q2B_VsB} the field dependences of $B^2Q$  of the different haloscopes are shown at 4~K and 9.08~GHz. 
Both a lower worst-case limit, obtained considering fluxons aligned in the $ab$ planes (for the anisotropic SC) and $\bm B\perp\bm J$, and an upper limit, obtained with the $\alpha$ coefficients evaluated as shown in Section~\ref{sec:Phys}, are given. It can be seen that in the worst case both Nb$_3$Sn and NbTi exhibit a lower $B^2Q$ than that given by a Cu cavity, while both YBCO and FeSeTe have, even in this case case, a  $B^2Q$ \red{higher than } that of Cu. In \textout{a} \red{the } more realistic scenario \textout{, shown through} \red{given by } the upper limits \textout{given} \red{shown } in Figure~\ref{fig:Q2B_VsB}, all the \red{SC here } studied \textout{SC} are better than Cu.
\textout{In particular at 5 T the numerical performances comparison is shown in Table~ref tab:B2 . In particular, }
\red{Table~\ref{tab:B2Q} reports the numerical evaluation of the performances at 5 T, where it can be seen that }
$(B^2Q)_{\mbox{{\scriptsize YBCO}}}/(B^2Q)_{\mbox{{\scriptsize Cu}}}$ is close to the upper limit of 14 reachable with the analysed geometry. At the highest $B$, NbTi performance starts dropping due its lower $B_{c2}$.
\red{With respect to YBCO, } FeSeTe \textout{despite its lower $\eta$ and $k_p$ with respect to YBCO,} exhibits similar performances \textout{to those of the latter} \red{despite its lower $\eta$ and $k_p$, because its } \textout{thanks to the} large $\lambda$ \textout{of FeSeTe which, in the bulk limit,} helps \textout{in reducing} \red{to reduce } $R_s$ \red{in the bulk limit, } as it can be deduced from Equation~\eqref{eqn:ZbulkSup}. 

Since at fixed signal to noise ratio SNR, bandwidth detection and system temperature, the integration time is $t_i\propto P_{a\rightarrow\gamma_{ph}}^{-2}$, then the increase of $(B^2Q)$ shown in Table~\ref{tab:B2Q} can be directly read as an equivalent reduction in $t_i$ \cite{braine2020extended}.

\begin{figure}[htbp]
\centering
\includegraphics[width=0.45\textwidth]{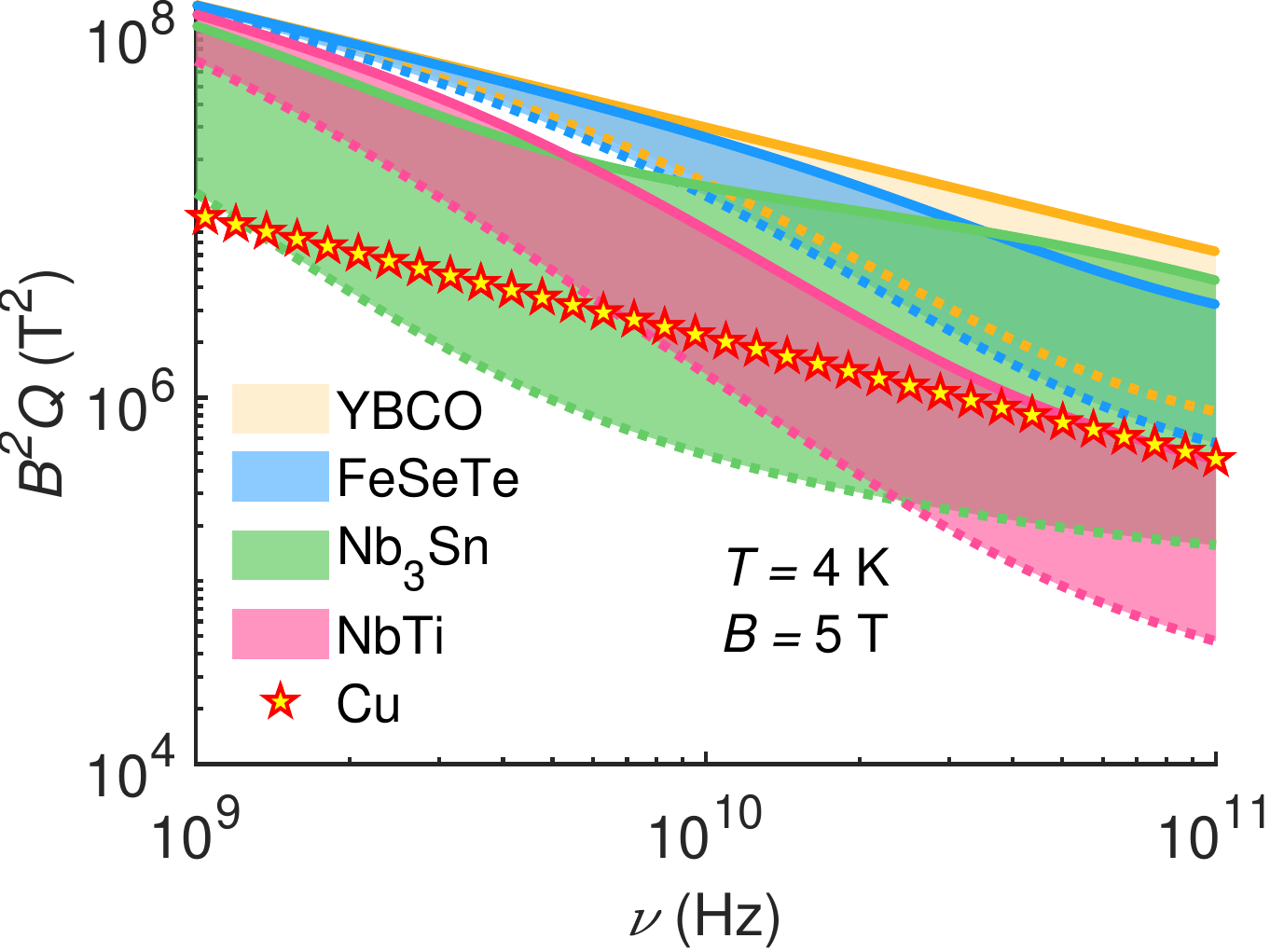}
\caption{Comparison of the $B^2Q$ factors evaluated at 4~K and 5~T as a function of $\nu$ and keeping the same geometrical factors \textout{shown} \red{as } in Section~\ref{sec:Haloscope}. The dotted lines represent the worst lower limit obtained in the $\bm B\perp\bm J$ configuration. The continuous lines represent a more realistic situation for which the $\alpha$ angular coefficient is taken into account as explained in the text. The ``star'' symbols represent the $B^2Q$ values obtained with a Cu cavity, the yellow area the values obtainable with YBCO, the light blue area those with FeSeTe, the green with Nb$_3$Sn and the red with NbTi.}
\label{fig:Q2B_VsF}
\end{figure}

The frequency dependence $B^2Q(\nu)$, reported in Figure~\ref{fig:Q2B_VsF} for ${10^9<\nu/\mbox{(Hz)}<10^{11}}$ at 5~T and 4~K,
shows interesting features that can be used to assess the best material in the different bands. 
It can be seen that, given the low $\nu_p$ of Nb$_3$Sn, the $Q$ of the associated haloscope is expected to decrease \red{at frequencies lower than in } \textout{before that of} the other materials. However, thanks to the low $\rho_n$ and high $B_{c2}$ of Nb$_3$Sn, with respect to \textout{those of} NbTi the figure of merit of the Nb$_3$Sn haloscope tends to settle on higher values at high frequency where $\tilde\rho_v$ saturates at $\rho_{ff}$. Thus, Nb$_3$Sn is more convenient than NbTi for $\nu>10$~GHz and, at even higher frequencies, its performances could be comparable with those of YBCO and FeSeTe. This result highlights clearly that for high frequency applications of SC in the mixed state, not only a high $\nu_p$ is desirable but also a low $\rho_{ff}$ is important.

 \begin{center}
 \begin{table}[htbp]
 \caption{Comparison of the $(B^2Q)$ figure of merit of the different materials with respect to that of Cu at 5~T, 4~K, 9.08~GHz.}\label{tab:B2Q}
 \centering
\begin{tabular}{|c|c|c|} 
 \hline
  Material & $(B^2Q)/(B^2Q)_{\mbox{{\scriptsize Cu}}}$& $t_i/t_{i,{\rm{Cu}}}$ \\ 
 \hline\hline
 NbTi & 4.3 & 0.05  \\
    \hline
     Nb$_3$Sn & 6.5 & 0.02 \\
    \hline
     FeSeTe& 12 & 0.006 \\
    \hline
     YBCO& 14 & 0.005  \\
   \hline
 \end{tabular}
 \end{table}
\end{center}

For what concerns the frequency dependence of $B^2Q(\nu)$ for FeSeTe and YBCO, it can be noted that the worst case lower limits of $B^2Q$ follow the same frequency behaviour due to their similar $\nu_p$. \textout{while} The upper $B^2Q$ limits follow a $\nu^{2/3}$ dependence (typical of the anomalous skin depth regime exhibited by Cu) for $\nu<10$~GHz for both YBCO and FeSeTe. In fact, with these materials, at the lower frequencies \red{the cavity losses } are dominated by the copper bases. At higher frequencies, the losses in YBCO still do not contribute substantially to the overall $Q$ while with FeSeTe $Q$ starts decreasing more rapidly than in the Cu cavity.

\section{Conclusion}
In this paper, the impact of \textout{the} vortex motion \textout{properties} and screening properties of different superconducting materials (i.e. NbTi, Nb$_3$Sn, YBCO and FeSeTe) on the performances of haloscopes, in terms of the power of the detected signal, was evaluated. 

In particular, the comparison of the $B^2Q$ figure of merit of haloscopes in the hybrid geometry proposed in \cite{alesini2019galactic} has been done by evaluating the $Q(B,\nu)$ as determined by the surface resistance $R_s(B,\nu)$ behaviour of the aforementioned SC.

In turn, $R_s$ of these SC was computed starting from the vortex motion parameters measured at microwave frequencies and high magnetic fields. 
\textout{, and by modelling t}
The effect on fluxon pinning by the different orientations of the static magnetic field $\bm B$ with respect to the induced microwave currents imposed by the specific haloscope geometry \red{was taken into account}.  

The results show that even at 5~T at low frequencies $\nu<10$~GHz all the studied materials allow to increase the haloscope sensitivity \red{with respect to Cu } while, at higher frequencies, NbTi becomes unsuitable and Nb$_3$Sn, despite its lower $\nu_p$, can still be \green{a useful } solution. Interestingly, FeSeTe, which \textout{is} here \red{is } for the first time considered for this application, despite its high vortex motion dissipation can be a promising material \textout{to be used} for the inner coating of haloscopes. Indeed, its large $\lambda$ contributes \textout{in lowering} \red{to lower } the bulk $R_s$ allowing to reach \textout{with this material comparable} performances \textout{of that} \red{comparable to those } obtained with YBCO. \textout{In addition to this} \red{Moreover, } one has to consider that Iron based SC are still far from the optimization and maturity levels reached with YBCO, thus allowing further room for improvements. 

Finally, YBCO is shown to be the best choice, \textout{allowing} \red{leading to } an increase \textout{on} \red{of } the measurement sensitivity by a factor $\sim13$  with respect to the Cu cavity in the whole frequency range analysed (i.e. ${10^9<\nu/\mbox{(Hz)}<10^{11}}$). This result is close to the maximum theoretical increase in the sensitivity reachable with the haloscope geometry studied.

To conclude, despite the expected high performances of YBCO in the operative conditions of haloscopes, this study shows the possibility \textout{of employing} \red{to employ } also other SC, in selected frequency bands, to increase the sensitivity of these axion detectors.

\vspace{6pt} 




\section*{Funding}
Work partially supported by MIUR-PRIN project ``HIBiSCUS'' - grant no. 201785KWLE and by the INFN- Commissione Scientifica Nazionale 5 project ``SAMARA''.

\section*{Acknowledgments}
The authors acknowledge C. Pira for the \textout{realization of the depositing NbTi thin films process on} \red{NbTi coating of } copper cavities, T. Spina and R. Fl\"ukiger for the Nb$_3$Sn sample, V. Pinto and G. Celentano for the YBCO sample, G. Sylva and V. Braccini for the FeSeTe sample.






\section*{References}
\providecommand{\newblock}{}

%


\end{document}